\documentclass[twocolumn,nofootinbib,amsmath,amssymb,aps,prd,balancelastpage,natbib]{revtex4-1}

\usepackage{graphicx}
\usepackage[caption=false]{subfig}

\usepackage{amsmath,amssymb}
\usepackage{amsfonts,amssymb,mathrsfs}

\usepackage[bookmarks=false,pdfstartview=FitH]{hyperref}
\usepackage[all]{hypcap}

\def\be{\begin{equation}}
\def\ee{\end{equation}}
\def\nn{\nonumber}
\def\f{\frac}

\def\sgn{{\rm sgn}}
\def\pl{{\rm Pl}}
\def\lp{\ell_\pl}
\def\b{\bar}
\def\d{\dot}
\def\h{\hat}
\def\t{\tilde}

\def\wh{\widehat}
\def\wt{\widetilde}

\def\dd{{\rm d}}

\def\de{\delta}
\def\ep{\epsilon}
\def\ga{\gamma}
\def\la{\lambda}
\def\om{\omega}
\def\ve{\varepsilon}

\def\mH{\mathcal{H}}
\def\mO{\mathcal{O}}
\def\mP{\mathcal{P}}
\def\mC{\mathcal{C}}
\def\mN{\mathcal{N}}
\def\oe{\mathring{e}}
\def\ow{\mathring{\omega}}
\def\oq{\mathring{q}}

\def\omu{\mathring{\mu}}
\def\lo{\ell_o}
\def\bra{\langle}
\def\ket{\rangle}
\def\tr{{\rm Tr}}
\def\cR{c_R}
\def\cI{c_I}
\def\mfu{\mathfrak{u}}

\usepackage{color}

\begin{document}

\pagestyle{plain}

\title{Loop quantum cosmology with self-dual variables}

\author{Edward Wilson-Ewing} \email{wilson-ewing@aei.mpg.de}
\affiliation{Max Planck Institute for Gravitational Physics (Albert Einstein Institute),\\
Am M\"uhlenberg 1, 14476 Golm, Germany, EU}

\begin{abstract}

Using the complex-valued self-dual connection variables, the loop quantum cosmology of a closed Friedmann universe coupled to a massless scalar field is studied.  It is shown how the reality conditions can be imposed in the quantum theory by choosing a particular inner product for the kinematical Hilbert space.  While holonomies of the self-dual Ashtekar connection are not well-defined in the kinematical Hilbert space, it is possible to introduce a family of generalized holonomy-like operators of which some are well-defined; these operators in turn are used in the definition of the Hamiltonian constraint operator where the scalar field can be used as a relational clock.  The resulting quantum theory is closely related, although not identical, to standard loop quantum cosmology constructed from the Ashtekar-Barbero variables with a real Immirzi parameter.  Effective Friedmann equations are derived, which provide a good approximation to the full quantum dynamics for sharply-peaked states whose volume remains much larger than the Planck volume, and they show that for these states quantum gravity effects resolve the big-bang and big-crunch singularities and replace them by a non-singular bounce.  Finally, the loop quantization in self-dual variables of a flat Friedmann space-time is recovered in the limit of zero spatial curvature and is identical to the standard loop quantization in terms of the real-valued Ashtekar-Barbero variables.

\end{abstract}


\maketitle

\section{Introduction}
\label{s.intro}

One of the key developments in loop quantum gravity (LQG) was the introduction of the complex-valued self-dual Ashtekar connection variables \cite{Ashtekar:1986yd, Ashtekar:1987gu}.  These variables played an important role as they not only made it possible to rewrite general relativity as a (constrained) gauge theory and thus suggested the use of holonomies in the quantum theory \cite{Jacobson:1987qk, Rovelli:1989za}, but also because the form of the constraints is greatly simplified when expressed in terms of the self-dual variables.

However, since the self-dual variables are complex-valued, it is necessary to impose reality conditions in order to recover general relativity.  The imposition of these reality conditions for generic space-times in the quantum theory has turned out to be very difficult to implement, and remains an open problem.  Furthermore, it is not known how to define a measure in the space of generalized connections on non-compact groups.  It is precisely these difficulties that have motivated the current use of the real-valued Ashtekar-Barbero connection variables \cite{Barbero:1994ap} in loop quantum gravity.

Indeed, the use of real-valued variables obviates the need of any reality conditions and, since the Ashtekar-Barbero connection is $\mathfrak{su(2)}$-valued, it is possible to construct the Ashtekar-Lewandowski measure in the space of generalized connections \cite{Ashtekar:1993wf}.  While these are important advantages, there do exist some drawbacks related to the use of the Ashtekar-Barbero variables.  First, it is necessary to introduce a new parameter in the theory, the Barbero-Immirzi parameter $\gamma$ \cite{Immirzi:1996di} (with $\gamma = \pm i$ for the original self-dual variables).  Second, the form of the scalar constraint becomes significantly more complicated due to the appearance of an additional term related to the spatial curvature.  While this new term can be handled in the quantum theory \cite{Thiemann:1996aw}, a number of supplementary quantization ambiguities associated to the presence of this additional term arise.  Finally, unlike the self-dual connection, the real-valued connection does not admit an interpretation as a space-time gauge field: it only tranforms as a connection under diffeomorphisms that are tangential to the spatial surface \cite{Samuel:2000ue, Alexandrov:2001wt}.
(An alternative way to render four-dimensional Lorentz covariance explicit, rather than using a space-time connection, is via projected spin networks where the SU(2) spin networks are viewed as projected SL(2,$\mathbb{C}$) spin networks \cite{Livine:2002ak}.  A number of the recent proposals for spin foam models can be expressed in this framework \cite{Dupuis:2010jn, Baratin:2011hp}.)

Furthermore, there are a number of recent papers studying black holes in loop quantum gravity that point out that a black hole entropy of $S = A / 4 \lp^2$ is naturally obtained after performing an analytic continuation sending $\gamma \to i$ \cite{Frodden:2012dq, Bodendorfer:2012qy, Han:2014xna, Achour:2014eqa}.  These results indicate that it may potentially be important to set the Barbero-Immirzi parameter to $\gamma = i$ in LQG.  Further discussions on this topic as well as some additional results pointing in the same direction can be found in \cite{Bodendorfer:2013hla, Pranzetti:2013lma, Geiller:2014eza, Carlip:2014bfa}.

The recent results in the studies of black holes, together with the drawbacks outlined above that are associated with the use of the real-valued connection, motivate a reexamination of the possibility of using the self-dual variables.  A first step in this direction is to consider highly symmetric space-times where the reality conditions have a relatively simple form, and the measure problem can often be completely avoided.  While the quantization of vacuum spherically symmetric and asymptotically flat space-times with self-dual variables has previously been studied in some detail \cite{Thiemann:1992jj}, there do not exist analogous studies of any other highly symmetric space-times where the quantization procedure becomes simpler, like homogeneous cosmological space-times.  One of the main goals of this paper is to explicitly show how the reality conditions can be imposed for the Friedmann-Lema\^itre-Robertson-Walker (FLRW) space-time.  While the presence of the symmetries of the FLRW space-times will simplify this task, the hope is that this result may give some insight into how the reality conditions can be applied in a more general context.  As already mentioned, it is harder to address the measure problem in this context since in minisuperspace models the problem of the non-compactness of the measure effectively vanishes.

Indeed, the imposition of the reality conditions will be one of the key steps in the study of the loop quantum cosmology (LQC) of the spatially closed FLRW space-time in terms of the self-dual variables that is presented in this paper.  Since there exists a vast literature in the field of loop quantum cosmology (for recent reviews, see e.g.\ \cite{Bojowald:2008zzb, Ashtekar:2011ni, Banerjee:2011qu}), many of the details of the quantization procedure ---when following steps analogous to those already well known in loop quantum cosmology--- will be only briefly described in this paper in order to avoid unnecessary repetition and the reader is referred to the original papers studying the `improved dynamics' LQC of spatially closed FLRW space-times in terms of the real-valued Ashtekar-Barbero connection \cite{Ashtekar:2006es, Szulc:2006ep}, which I will refer to as `standard LQC' from now on.  Rather, the focus of this paper will be on the differences that arise when one works with self-dual variables.

The outline of the paper is as follows: there is a brief introduction to the self-dual Ashtekar variables in Sec.~\ref{s.var}, which is followed by a description in Sec.~\ref{s.cosm} of the symmetry reduction necessary in order to study the spatially closed FLRW space-time in terms of these variables.  Then in Sec.~\ref{s.qc} the quantum theory is presented in two steps ---the first being a definition of the kinematical Hilbert space where following the procedure suggested in \cite{Ashtekar} the reality conditions are imposed through a careful choice of the inner product, and the second being the construction of the Hamiltonian constraint operator, with the important result that the classical big-bang and big-crunch singularities are resolved--- and in the last part of this section the resulting self-dual version of LQC is compared to standard LQC.  The effective equations for the quantum theory are studied in Sec.~\ref{s.eff}, and there is a discussion in Sec.~\ref{s.disc} which includes a comparison of the results obtained in this paper with those of \cite{Achour:2014rja} where an analytic continuation is used in order to define a different version of LQC with a Barbero-Immirzi parameter of $\gamma = i$.

\section{Self-Dual Variables}
\label{s.var}

This section contains a very brief review of the complex-valued self-dual variables that will be used in this paper.  The review is restricted to the definitions and results that will be needed in the following sections and is thus necessarily incomplete.  For further details, see e.g.\ \cite{Ashtekar} and the many references therein.

The self-dual variables are the Ashtekar connection $A_a^k$ and the densitized triads $E^a_k$, which are related to the triads $e^a_k$, the determinant of the spatial metric $q$, the spin-connection $\Gamma_a^k$ and the extrinsic curvature $K_a^k = K_{ab} e^{bk}$ as follows:
\be 
A_a^k = \Gamma_a^k + i K_a^k, \qquad E^a_k = \sqrt{q} \, e^a_k.
\ee

These variables are canonically conjugate,
\be 
\{A_a^j(x), E^b_k(y)\} = i \cdot 8 \pi G \, \de_a^b \, \de^j_k \, \de^{(3)}(x-y),
\ee
and one of their main advantages is that the scalar constraint takes a simple form,
\be
\mH = \f{E^a_i E^b_j}{16 \pi G \sqrt{q}} \epsilon^{ij}{}_k F_{ab}{}^k + \f{\pi_\phi^2}{2 \, \sqrt{q}} \approx 0,
\ee
with the field strength of the self-dual connection being
\be
F_{ab}{}^k = 2 \, \partial_{[a} A_{b]}^k + \epsilon_{ij}{}^k A_a^i A_b^j,
\ee
and here a massless scalar field is taken as the matter content, with $\pi_\phi = \sqrt{q} \, \partial_t \phi$.  Note that the spatial curvature term $(1+\ga^2) \Omega_{ab}{}^k$ that appears in the scalar constraint for Ashtekar-Barbero variables vanishes since $\ga = i$.  The Gauss and diffeomorphism constraints also have a similarly simple expression \cite{Ashtekar}.

Since the variables $A_a^k$ and $E^a_k$ are complex-valued (and the spatial metric and extrinsic curvature appearing in the Arnowitt-Deser-Misner Hamiltonian formulation of general relativity are not), it is necessary to impose reality conditions which take the form
\be 
A_a^k + \left(A_a^k\right)^\star = 2 \, \Gamma_a^k, \qquad
E^a_k E^{bk} > 0.
\ee
As mentioned in the Introduction, it is currently not known how to correctly implement these reality conditions in the full quantum theory and this is one of the main reasons the real-valued Ashtekar-Barbero connection is now typically used in loop quantum gravity.

The fundamental variables in the quantum theory will correspond to (i) what I will call `generalized holonomies' along (oriented) edges $e$
\be \label{def-gh}
h_e = \mP \exp \left[ \int_e  \alpha A_a \right],
\ee
with $A_a := A_a^i \, \sigma_i / 2i$, where the $\sigma_i$ are the Pauli matrices and $\alpha$ is a complex-valued parameter, and (ii) areas of surfaces.  For standard holonomies $\alpha = 1$, but as shall be shown below these holonomies are not well-defined in the kinematical Hilbert space of self-dual loop quantum cosmology given in Sec.~\ref{ss.kin}.  Instead, it is necessary that $\alpha$ be purely imaginary for the operators corresponding to generalized holonomies to be well-defined in self-dual LQC.

Generalized holonomies contain the same geometric information as standard holonomies (albeit in a rescaled form), and they have three of the same important properties:
\begin{enumerate}
\item the generalized holonomy of an $\mathfrak{sl(2, C)}$-valued connection is an $SL(2, \mathbb{C})$ group element,
\item the composition rule for generalized holonomies is the usual one, $h_{e_1 \circ e_2} = h_{e_1} h_{e_2}$,
\item the generalized holonomy along the inverted edge is the inverse, $h_{e^{-1}} = h_e^{-1}$.
\end{enumerate}
Furthermore, the Poisson algebra of generalized holonomies and fluxes will be very similar to the standard holonomy-flux algebra with the only difference being the presence of an additional factor of $\alpha$ in the result of the Poisson bracket.  On the other hand, under gauge transformations generalized holonomies do not transform in as simple a fashion as standard holonomies.

There is also another important difference between standard and generalized holonomies.  Defining
\be
F_{ab} = 2 \partial_{[a} A_{b]} + [A_a, A_b],
\ee
and
\be
F_{ab}{}^k = 2 \partial_{[a} A_{b]}^k + \ep_{ij}{}^k A_a^i A_b^j,
\ee
the relation
\be
F_{ab} = \f{\t\alpha}{2i} F_{ab}{}^k \, \sigma_k
\ee
holds if and only if $\alpha = 1$ (in which case $\t\alpha = 1$ also).  For this reason, if $\alpha$ is left free, it is important to remember that while generalized holonomies of $A_a$ will be related in a simple fashion to $A_a^i$,
\be \label{a-hol}
A_a^i \cdot t^a = \lim_{z\to0} \f{i \, \tr [ (\mP e^{\, \int_0^z \! \alpha A_a t^a}) \sigma^i]}{\alpha z},
\ee
(where $t^a$ is the tangent vector to the path and $z$ is the length of the path), the relation between the field strength $F_{ab}$ obtained from the generalized holonomy of $A_a$ around a closed loop and the quantity $F_{ab}{}^k$ which appears in the scalar constraint is considerably more complicated.  Because of this, in Sec.~\ref{ss.ham} it will be more convenient to express the scalar constraint operator in self-dual loop quantum cosmology in terms of a non-local connection operator (constructed from generalized holonomies) rather than a non-local field strength operator as is commonly done for standard isotropic loop quantum cosmology.  Note that it is also necessary to introduce a non-local connection operator in order to properly define the Hamiltonian constraint operator for Bianchi space-times with non-vanishing spatial curvature \cite{Ashtekar:2009um, WilsonEwing:2010rh}, and this connection-based loop quantization has since also been studied for FLRW space-times in standard LQC \cite{Corichi:2011pg, Corichi:2013usa, Singh:2013ava}.

\section{Cosmology}
\label{s.cosm}

The metric of a closed FLRW space-time is
\be \label{metric}
ds^2 = -dt^2 + a(t)^2 \ow_a^i \ow_b^j \delta_{ij} dx^a dx^b,
\ee
where the fiducial co-triads $\ow_a^i$ satisfy the relation
\be \label{fid}
\dd \ow^i + \f{1}{2} \mathring\epsilon^i{}_{jk} \ow^j \wedge \ow^k = 0,
\ee
with $\mathring{\epsilon}_{ijk}$ being totally anti-symmetric and $\mathring{\epsilon}_{123}=1$.  The fiducial triads $\oe^a_i$ (the inverse of the co-triads $\ow_a^i$) provide a basis to a three-sphere of radius 2 and volume $V_o = 16 \pi^2$.

The spatial part of the metric \eqref{metric} can be rewritten in terms of co-triads as $\omega_a^k = a(t) \, \ow_a^k$ and from this and \eqref{fid}, it follows that the spin-connection has the simple form
\be
\Gamma_a^k = -\epsilon^{ijk} e^b_j \left( \partial_{[a} \om_{b]k} + \f{1}{2}
e^c_k \om_a^l \partial_{[c} \om_{b]l} \right)
= \f{\ve}{2} \, \ow_a^k,
\ee
with $\epsilon^{123} = \ve = \pm1$, where the sign is determined by the handedness of the triads.  (The antisymmetric tensor $\epsilon_{ijk} = e^a_i e^b_j e^c_k \epsilon_{abc}$ is related to the volume 3-form of the spatial surface and is not to be confused with the structure constants $\mathring\epsilon_{ijk}$ that appear in \eqref{fid}.  For a detailed discussion concerning the relation between the orientation of the triads and the sign of $\epsilon^{123}$, see Sec.~IIA in \cite{Ashtekar:2009um}).

A convenient parametrization of the self-dual Ashtekar connection and the densitized triads is
\be \label{def-AE}
A_a^k = \f{c}{\lo} \, \ow_a^k, \qquad E^a_k = \f{p}{\lo^2} \, \sqrt{\oq} \, \oe^a_k,
\ee
where $\oq$ is the determinant of the fiducial metric $\oq_{ab} = \ow_a^i \ow_{bi}$ and $\lo := V_o^{1/3}$.  In classical general relativity, it is easy to show that $p = \sgn(a) a^2 \lo^2$ and $c = \lo[\ve/2 + i \dot{a}]$, where the dot indicates a derivative with respect to the proper time $t$. Also, note that $p$ (and $a$) can be negative as they encode the orientation of the triads: a positive $p$ corrresponds to right-handed triads and a negative $p$ to left-handed triads.  Therefore, $\ve = \sgn(p)$.  Since the handedness of the triads does not affect the dynamics of the space-time, neither does the sign of $p$.  For further details on this parametrization of a closed FLRW space-time, see \cite{Ashtekar:2006es, Bojowald:1999tr}.

This parametrization induces from the full theory the symplectic structure
\be \label{poisson}
\{c, p\} = i \cdot \f{8 \pi G}{3},
\ee
and the reality conditions reduce to
\be \label{real-cl}
c + c^\star = \ve \, \lo, \qquad p^\star = p.
\ee

Finally, it is easy to check that the parametrization \eqref{def-AE} automatically satisfies the diffeomorphism and Gauss constraints, so the Hamiltonian constraint is simply given by $\mC_H = \int N \mH$ (with $N$ being the lapse),
\be \label{ham-cl}
\mC_H = \f{3 N \sqrt{|p|} \, \ve^2}{8 \pi G} \Big[ \ve^2 c^2 - \ve \, \lo c \Big] + \f{N p_\phi^2}{2 |p|^{3/2}} \approx 0.
\ee
The momentum of the scalar field is $\pi_\phi = |p|^{3/2} \d \phi \sqrt{\oq} / V_o$, and the variable $p_\phi = |p|^{3/2} \d \phi$ is introduced in order to simplify the notation.  This is a convenient choice since the resulting Poisson bracket has a simple form, $\{\phi, p_\phi\} = 1$.  There are a number of $\ve = \sgn(p)$ terms in the constraint; the first $\ve^2$ arises from rewriting $p^2 = |p|^2 \ve^2$, and the others come from $\epsilon_{123} = \ve$.  In the classical theory it is possible to choose a particular orientation and set, for example, $\ve = \sgn(p) = 1$, but in the quantum theory it is important to be able to consider superpositions of states with positive and negative $p$.  While it is tempting to set $\ve^2=1$, the presence of one such $\ve^2$ term will ultimately permit a factor-ordering choice that simplifies the action of the Hamiltonian constraint operator in the quantum theory.  Therefore, no $\ve^2$ terms will be set to 1 at this stage.

Note that for the flat FLRW space-time, \eqref{fid} becomes simply $\dd \ow^i = 0$, which leads to a vanishing spin-connection.  Thus, a convenient shortcut in order to obtain the reality conditions and the Hamiltonian constraint for the flat FLRW cosmology is to set $\lo = 0$ in \eqref{real-cl} and \eqref{ham-cl} (this shortcut is also used in standard LQC in order to obtain the Hamiltonian constraint for the spatially flat case coming from the spatially closed FLRW space-time \cite{Ashtekar:2006es}).  This is what can be done at the end of this paper in order to extend the results obtained here for the case of vanishing spatial curvature.

The classical Friedmann equation is obtained by squaring the relation
\be
\dot p = \{p, \mC_H\} = -i \cdot 3 \sqrt{|p|} (2 c - \lo),
\ee
(for the lapse $N=1$ and having set $\ve=1$), and using the constraint $\mC_H = 0$ in order to obtain
\be \label{fr-cl}
H^2 = \f{\dot p^2}{4 p^2} = \f{8 \pi G}{3} \rho - \f{\lo^2}{4 |p|}.
\ee
Here $H = \dot a / a$ is the Hubble rate and $\rho = p_\phi^2 / 2 |p|^3$ is the energy density of the massless scalar field.  Note that the factor of 4 in the denominator of the second term appears due to the conventions in \eqref{fid} where the radius of the three-sphere is taken to be 2.

Together with the continuity equation
\be \label{cont}
\dot \rho + 3H (\rho + P) = 0,
\ee
where $P$ is the pressure of the matter field, the Friedmann equation determines the dynamics of the space-time.  For a massless scalar field, $P = \rho$ and the continuity equation simplifies to $\d p_\phi = 0$.

An important point here is that there is no need to dynamically impose reality conditions if one is working with the equations \eqref{fr-cl} and \eqref{cont}: if the scale factor and energy density are initially taken to be real, their reality will be preserved by the dynamics generated by \eqref{ham-cl}.

Finally, as explained above in Sec.~\ref{s.var}, generalized holonomies of the Ashtekar connection will play a key role in the quantum theory.  An important result following from \eqref{def-gh} is that the generalized holonomy of $A_a^k$ taken along a path tangential to the fiducial triad $\oe_k$ and of a length $\omu \lo$ with respect to the fiducial metric is
\begin{align}
h_k(\omu) &= \exp \left( \f{\alpha \omu c \sigma_k}{2i} \right) \nn \\ &
\label{hol}
= \cosh \left( \f{\alpha \omu c}{2i} \right) \mathbb{I} + \sinh \left( \f{\alpha \omu c}{2i} \right) \sigma_k.
\end{align}

\section{Quantum Cosmology}
\label{s.qc}

This section is divided in three parts: the first contains the definition of the kinematical Hilbert space of the theory ---including the implementation of the reality conditions through the choice of the inner product--- while the second defines the Hamiltonian constraint operator for the theory and makes explicit its action on generic states, and the third compares self-dual LQC to standard LQC.

\subsection{The Kinematical Hilbert Space}
\label{ss.kin}

The kinematical Hilbert space of the theory has two sectors, namely the gravitational and matter sectors, and the total Hilbert space is simply given by their tensor product,
\be
H_{\rm tot} = H_g \otimes H_m,
\ee
with wave functions living in each of these Hilbert spaces denoted by $\psi(c)$ and $\chi(\phi)$ respectively.

Starting with the gravitational sector, following loop quantum gravity the fundamental operators in the quantum theory are taken to be generalized holonomies of $A_a^k$ and areas of surfaces.  Due to homogeneity, it is enough to consider generalized holonomies that are parallel to the fiducial triads $\oe^a_k$ [in which case the $c$-dependence only appears in sums and differences of terms of the form $\exp(\alpha \omu c / 2i)$ as can be seen in \eqref{hol}], and simply define an operator corresponding to the classical variable $p$ for an area operator.

Since the self-dual variables are complex, the wave functions $\psi(c)$ are required to be holomorphic functions of $c$.  Then, following the standard procedure of LQC, a natural choice for the definition of the fundamental operators is
\be \label{basic-ops}
\hat p \, \psi(c) = \f{8 \pi G \hbar}{3} \f{d}{dc} \, \psi(c), \qquad
\wh{e^{\mu c}} \, \psi(c) = e^{\mu c} \, \psi(c),
\ee
where $\mu = \alpha \omu / 2i$.  As explained above, since linear combinations of $\wh{e^{\mu c}}$ encode the non-trivial information of a generalized holonomy of length $\omu \lo$ (with respect to the fiducial metric $\oq_{ab}$) along an edge parallel to any one of the $\oe^a_k$, and $\h p$ is proportional to the equatorial area of the 3-sphere, these operators are indeed the LQC equivalent of the fundamental LQG operators, namely (generalized) holonomies and areas.

In terms of these basic operators, the reality conditions are
\be \label{real}
\h p^\dag = \h p, \qquad \left( \wh{e^{\mu c}} \right)^\dag = \wh{ \left(e^{-\bar\mu \, [c - \sgn(p) \lo]} \right)},
\ee
which, following \cite{Ashtekar}, shall be implemented by making a specific choice for the inner product.  Note that since $\sgn(p)$ appears in the exponential on the right-hand side of the second operator equation, there is a factor-ordering ambiguity.  At this point, the operator on the right-hand side is not completely defined since no factor-ordering choices have been made yet.  However, as shall be shown below, a preferred factor-ordering will be selected for this operator by the inner product that is constructed in the following paragraphs, and this factor-ordering choice will complete the definition of the second reality condition.

The eigenstates of $\hat p$ are
\be \label{def-p}
|p\ket = e^{3 p c / 8 \pi G \hbar},
\ee
and while a priori $p \in \mathbb{C}$, the reality condition $\h p^\dag = \h p$ implies that writing $|p\ket$ ---which suggests that this state is normalizable (as it shall be shown to be below for real-valued $p$)--- only makes sense for $p \in \mathbb{R}$.

The other basic operator of LQC acts on these eigenstates as a shift operator,
\be \label{shift-op}
\wh{e^{\mu c}} \, |p\ket = |p + 8 \pi G \hbar \mu / 3 \ket.
\ee
Immediately from this result it can be seen that it is necessary to demand that $\alpha$ be purely imaginary since the reality conditions impose that $p$ be real.  Therefore, assuming that $p$ is initially real, for the shifted state to also represent a real-valued area, $\mu = \alpha \omu / 2i$ must be real.  Since $\omu$ is by definition real, this condition requires that $\alpha$ be purely imaginary.  Furthermore, as shall be shown below, $|p\ket$ is normalizable if and only if $p \in \mathbb{R}$.  Therefore, it follows that for the action of the shift operator to be well-defined on $H_g$, it is necessary that $\mu$ be real-valued.  This condition rules out operators corresponding to standard holonomies and only allows generalized holonomies with $\alpha$ a purely imaginary number.  (While at present there is no reason to prefer a specific numerical value for $\alpha$, it seems reasonable to assume that it is of the order of unity, with $i$ appearing to be a somewhat natural choice since for standard holonomies $\alpha = 1$.)

The inner product must be defined so that it implements the reality conditions \eqref{real}.  In order to avoid complications related to the factor-ordering ambiguities, it is easiest to start by considering wave functions that only have support on positive $p$,
\be \label{psi+}
\psi^+(c) = \sum_{p>0} C_p |p\ket,
\ee
and require that the simplified reality conditions for positive $p$
\be \label{real+}
\h p^\dag = \h p, \qquad \left( \wh{e^{\mu c}} \right)^\dag = e^{\bar\mu \, \lo} \, \wh{e^{-\bar \mu c}},
\ee
are imposed by the inner product.  Now there is no factor-ordering ambiguity and it will be a relatively straightforward process to determine the inner product for states of the type $\psi^+(c)$.

In order to find the appropriate inner product, a good starting point is provided by the ansatz
\be \label{innerP+}
\bra \psi^+_1 | \psi^+_2 \ket = \!
\lim_{L\to\infty} \f{1}{2L} \int_{-L}^L \!\!\!\!\! d\cR \int_{-L}^L \!\!\!\!\! d\cI \,
\mfu^+\!(c, \b c) \, \bar\psi^+_1 \psi^+_2 \!,
\ee
which is motivated by the fact that it is the space of square-integrable functions with respect to the Haar measure on the Bohr compactification of the real line that is used as the kinematical Hilbert space in standard LQC.  Here $c = \cR + i \cI$, and the measure $\mfu^+(c, \b c)$ must be chosen so that the reality conditions \eqref{real+} are implemented for wave functions of the type $\psi^+(c)$.  

Note that the ansatz \eqref{innerP+} is simply an intermediate step which will provide guidance for the ultimate definition of the inner product.  In fact, this form will only be used for the simple case $\sgn(p)=1$ (and an analogous form can be defined for $\sgn(p)=-1$).  From these results, the appropriate definition of the inner product for $\psi(c)$ with support on any linear combination of $|p\ket$ will become clear, and it will not be necessary to write the inner product for generic $\psi(c)$ in the form of an integral over $\cR$ and $\cI$.

The first reality condition in \eqref{real+} ---after integrating by parts and using the holomorphicity of the wave function $\psi^+(c)$--- gives the result that $\mfu^+(c, \b c) = \mfu^+(c + \b c)$.  The second condition in \eqref{real+} implies that
\be \label{int1}
\bra e^{\mu c} \psi^+_1| \psi^+_2\ket = \int_c \, \mfu^+(c, \bar c) \, e^{\b\mu \b c} \bar\psi^+_1 \, \psi^+_2,
\ee
[where $\int_c$ is shorthand for the integrals over $\cR$ and $\cI$ given in \eqref{innerP+}, including the prefactor of $1/2L$ and the limit of $L \to \infty$] and
\be \label{int2}
\bra \psi^+_1 | e^{\b\mu(\lo - c)} \psi^+_2 \ket = \int_c \, \mfu^+(c, \bar c) \, \bar\psi^+_1 \, e^{\b\mu(\lo - c)} \psi^+_2,
\ee
must be equal.  At this stage, it is necessary to choose $\mu$ in the expressions above to be real and with a value such that the shifted wave functions remain of the type \eqref{psi+} since this was an assumption for writing \eqref{innerP+}.  This is not problematic since, no matter the state $\psi^+(c)$, there always exists a sufficiently small $\mu$ that satisfies this condition.  [The restriction on the amplitude of $\mu$ will be lifted below once the inner product is defined for all $\psi(c)$.  However, $\mu$ will always be required to be real for the operator \eqref{shift-op} to be well-defined, as explained in the paragraph following Eq.~\eqref{shift-op}.]  The requirement that the two integrals \eqref{int1} and \eqref{int2} be equal imposes that the measure $\mfu^+(c, \b c)$ be distributional and have the form
\be
\mfu^+(c, \b c) = \de(c + \b c - \lo),
\ee
where the overall numerical factor is set to 1 for simplicity.  Note that this also satisfies the requirement of the first reality condition that $\mfu^+(c, \b c) = \mfu^+(c + \b c)$.

With this measure, it is easy to check that for positive $p$ the eigenstates $|p\ket$ are normalizable and that the inner product between two eigenstates $|p_1\ket$ and $|p_2\ket$ (assuming $p_1$ and $p_2$ are real and positive) is proportional to the Kronecker delta,
\be \label{inner-b+}
\bra p^+_1 | p^+_2 \ket = e^{3 \lo p_1 / 8 \pi G \hbar} \, \de_{p_1, p_2}.
\ee

Exactly the same procedure can be followed for wave functions that only have support on negative $p$,
\be
\psi^-(c) = \sum_{p<0} C_p |p\ket,
\ee
in which case the reality conditions become
\be
\h p^\dag = \h p, \qquad \left( \wh{e^{\mu c}} \right)^\dag = e^{-\bar\mu \, \lo} \, \wh{e^{-\bar \mu c}},
\ee
and the final result is that, for $p_1, p_2$ real and negative,
\be \label{inner-b-}
\bra p^-_1 | p^-_2 \ket = e^{-3 \lo p_1 / 8 \pi G \hbar} \, \de_{p_1, p_2}.
\ee

Then, with the natural definitions that the inner product between states with support on positive $p$ and negative $p$ vanish,
\be
\bra \psi^+_1 | \psi^-_2 \ket = 0,
\ee
and that the eigenstate for $p=0$ has norm 1 [obtained by calculating the inner product using equivalently \eqref{inner-b+} or \eqref{inner-b-}] and is orthogonal to all other eigenstates $|p' \neq 0\ket$,
\be
\bra p' | p=0 \ket = \de_{p', 0},
\ee
the inner product for the eigenstates of $\h p$ with $p \in \mathbb{R}$ is then given by
\be \label{innerP-basis}
\bra p_1 | p_2 \ket = e^{3 \lo |p_1| / 8 \pi G \hbar} \, \de_{p_1, p_2}.
\ee
From the reality condition $\h p^\dag = \h p$ it follows that only real-valued eigenvalues of $\h p$ are normalizable, and therefore the basis \eqref{def-p} with real-valued $p$ spans $H_g$.

Given this inner product, it can be seen that any normalized wave function in $H_g$ can be written as a countable sum
\be \label{norm-psi}
|\psi\ket = \sum_{p \in \mathbb{R}} C_p \, |p\ket,
\ee
with $\sum_p |C_p|^2  \, e^{3 \lo |p| / 8 \pi G \hbar} = 1$.

Recall that in the second of the operator equations \eqref{real} there remained a factor-ordering ambiguity.  Now, given the inner product \eqref{innerP-basis}, there exists a preferred factor-ordering choice, namely
\be \label{real2}
\left( \wh{e^{\mu c}} \right)^\dag = \wh{e^{-3 |p| \lo / 8 \pi G \hbar}} \; \wh{e^{-\mu c}} \; \wh{e^{3 |p| \lo / 8 \pi G \hbar}},
\ee
that is picked out by this inner product.  (Note that $\b\mu=\mu$ since $\mu$ must be real-valued, as explained above earlier.)  It is easy to check that this operator equation appropriately encodes the classical reality conditions for the quantum theory.  An extension of the classical reality conditions is necessary for situations where the sign of $p$ may change (this being impossible in classical general relativity), and \eqref{real2} gives a clear prescription of how this case is to be handled in the quantum theory.  Thus, the operator equations $\h p^\dag = \h p$ and \eqref{real2} not only perfectly capture the classical reality conditions everywhere they were originally defined, but also automatically extend them in an appropriate fashion in the regime where it is necessary to do so.

As an aside, note that it is possible to define a family of normalized eigenfunctions of $\h p$,
\be
|p\ket_n = e^{-k |p|} \, |p\ket, \quad \Rightarrow \quad
{}_n \bra p_1|p_2\ket_n = \de_{p_1,p_2},
\ee
where $k = 3 \lo / 16 \pi G \hbar$, in which case it follows that
\begin{align}
\h p |p\ket_n &= p |p\ket_n, \\
\wh{e^{\mu c}}|p\ket_n &= e^{k (|p + 8 \pi G \hbar \mu / 3| - |p|)}
|p + 8 \pi G \hbar \mu / 3 \ket_n. \label{shift-n}
\end{align}
While the basis $|p\ket_n$ can be convenient for some calculations ---e.g.\ showing that the reality conditions \eqref{real} are satisfied--- and does have the nice property that%
\footnote{On the other hand, $\bra p | \h p | p \ket = p \, e^{2 k |p|}$.}
${}_n\bra p| \h p | p \ket_n = p$, the basis $|p\ket$ clearly behaves in a simpler fashion under the action of the shift operators (and therefore also under the action of the Hamiltonian constraint operator); for this reason $\psi(p)$ shall be expressed in the form \eqref{norm-psi} in Sec.~\ref{ss.ham}.

Finally, since the sign of $p$ classically only determines the orientation of the triads and does not affect the metric, the wave function $\psi(p)$ should be invariant under the parity transformation $p \to -p$.  The parity transformation acts on the basis kets as
\be
\Pi \, |p\ket = |-p\ket,
\ee
and then the requirement that the wave function be invariant under a parity transformation is simply
\be \label{parity}
\Pi \, \psi(p) = \psi(p).
\ee

This completes the definition of the kinematical Hilbert space of the gravitational sector of the theory.  While the kinematical Hilbert space of self-dual LQC is very similar to that of standard LQC, there is one important difference: the shift operator \eqref{shift-op} does not preserve the norm of a wave function.  This is an unavoidable consequence of imposing the reality conditions via the choice of the inner product.

For the matter sector, the kinematical Hilbert space $H_m$ corresponds to the space of square-integrable functions $\chi(\phi)$ over the real line with respect to the Lebesgue measure, and the fundamental operators correspond to the scalar field $\phi$ and its momentum $p_\phi$,
\be
\hat \phi \chi(\phi) = \phi \chi(\phi), \qquad
\hat p_\phi \chi(\phi) = -i \hbar \f{d \chi(\phi)}{d \phi},
\ee
which act by multiplication and differentiation respectively.

It is clear that in order to obtain the kinematical Hilbert space for a flat FLRW cosmology, it is sufficient to set $\lo = 0$ in all of the results in this section with the result that the reality conditions and the inner product on $H_g$ simplify and the resulting kinematical Hilbert space is identical to the kinematical Hilbert space of standard LQC for a flat FLRW space-time.

\subsection{The Hamiltonian Constraint Operator}
\label{ss.ham}

In order to determine the physical Hilbert space, it is necessary to construct the Hamiltonian constraint operator and find the states that it annihilates.  In the classical Hamiltonian constraint, the quantities $p$, $c$ and $p_\phi$ appear.  While operators corresponding to $\h p$ and $\h p_\phi$ exist, there is no operator directly corresponding to the connection.  Instead, it is necessary to use operators of the type \eqref{basic-ops}, and therefore to define a non-local connection operator expressed in terms of generalized holonomies.

This can be done by employing the relation \eqref{a-hol} and, instead of taking the limit of the path length going to zero (in which case the result would not be well-defined as an operator on the Hilbert space), the path length (with respect to the physical metric) is set to be a minimal length $\ell_m$, assumed to be of the order of the Planck length.  This choice is of course motivated by the Planck-scale discreteness of (self-dual) LQG \cite{Rovelli:1994ge}, which implies that it is impossible to take the generalized holonomy of the connection along a path that is shorter than the Planck length.

Note that a similar procedure was first used to define a non-local connection operator in standard LQC for Bianchi space-times with a non-vanishing spatial curvature \cite{Ashtekar:2009um, WilsonEwing:2010rh}, where it was impossible to define a non-local field strength operator as is usually done for isotropic or spatially flat space-times in standard LQC, and it has since also been used for FLRW space-times in standard LQC \cite{Corichi:2011pg, Corichi:2013usa}.

The task is then to choose an appropriate value for the length $\omu$ appearing in the definition of the generalized holonomy.  Recall from \eqref{hol} that the length of the generalized holonomy with respect to the fiducial metric $\oq_{ab}$ is $\omu \lo$ and that therefore the physical length is $\omu \lo |a(t)| = \omu \sqrt{|p|}$.  Since the physical length of the generalized holonomy has been chosen to be set to $\ell_m$, this gives $\omu = \ell_m / \sqrt{|p|}$, or equivalently
\be
\mu = \f{\la_m}{\sqrt{|p|}},
\ee
where $\la_m = \alpha \ell_m / 2i$.  This choice for $\mu$ corresponds to the `improved dynamics' of standard loop quantum cosmology, as first introduced in \cite{Ashtekar:2006wn}.  The choice of $\ell_m$ ---which gives the scale of the underlying discreteness of the quantum theory--- in standard LQC is set to be the square root of the minimal non-zero eigenvalue of the area operator of loop quantum gravity.  A similar procedure can be followed here using known results about the spectrum of the area operator for self-dual variables which would set $\ell_m^2 = 4 \sqrt{3} \pi G \hbar$ \cite{Rovelli:1994ge}, but in principle $\ell_m$ should be determined by a derivation of LQC from full LQG.

From this, it follows that the non-local connection operator has the form
\begin{align}
\h c &= \f{\sqrt{|p|}}{\la_m} \sinh \f{\la_m c}{\sqrt{|p|}} \nn \\ & \label{nl-conn}
= \f{\sqrt{|p|}}{2 \la_m} \left( e^{\la_m c / \sqrt{|p|}} - e^{-\la_m c / \sqrt{|p|}} \right),
\end{align}
where factor-ordering ambiguities have been ignored for now.  The final step, before choosing a factor-ordering and calculating the action of the Hamiltonian constraint, is to define how the operator $\exp(\la_m c / \sqrt{|p|})$ acts.  Since the prefactor to $c$ depends on $p$, the simple shift operation defined in \eqref{basic-ops} for the case of a constant prefactor cannot be applied here.

A solution is offered by the fact that the classical variable conjugate to $\beta = c/\sqrt{|p|}$ is $V = \sgn(p) |p|^{3/2}$,
\be
\{\beta, V\} = i \cdot 4 \pi G,
\ee
and therefore this operator should act as a simple shift operator on the eigenkets $|V\ket$ corresponding to the $\h V$ operator,
\be
e^{\la_m c / \sqrt{|p|}} | V \ket = | V + 4 \pi G \hbar \la_m \ket,
\ee
where the eigenvectors $|V\ket$ are simply a relabeling of the eigenvectors $|p\ket$ defined in \eqref{def-p} with $p = \sgn(V) |V|^{2/3}$.  Note that $\sgn(p) = \sgn(V)$ and therefore the sign of $V$ encodes the handedness of the triads.  Finally, in order to use a variable as similar to that of standard LQC as possible,
\be
\nu = \f{V}{2 \pi G \hbar \la_m}
\ee
is defined and then
\be
e^{\la_m c / \sqrt{|p|}} | \nu \ket = | \nu + 2 \ket.
\ee
The inner product on $H_g$ of these basis vectors is
\be \label{inner-nu}
\bra \nu_1 | \nu_2 \ket = e^{3 \lo \la_m^{2/3} |\nu|^{2/3} / 4 (2 \pi G \hbar)^{1/3}} \, \de_{\nu_1, \nu_2}.
\ee

As an aside, note that the action of $\exp(\la_m c / \sqrt{|p|})$ acts on the normalized eigenvectors $|\nu\ket_n$ as
\be
e^{\la_m c / \sqrt{|p|}} | \nu \ket_n = e^{\t k (|\nu+2|^{2/3} - |\nu|^{2/3})} | \nu + 2 \ket_n,
\ee
with $\t k = 3\lo \la_m^{2/3}/8(2 \pi G \hbar)^{1/3}$.  Since the action of this shift operator is very complicated when expressed in terms of the normalized eigenvectors, it is more convenient to use the non-normalized basis with the inner product \eqref{inner-nu} and this is what shall be done for the remainder of the paper.

The final step is to choose a lapse and a factor-ordering for the Hamiltonian constraint operator corresponding to $\mC_H$.  While the factor-ordering choices should ultimately be determined by the full theory of quantum gravity, at this time no particular choice is preferred, apart perhaps from aesthetical considerations.  There are obviously many possibilities, and a particularly convenient choice that simplifies the action of the constraint is to take the lapse to be $N = |p|^{3/2}$ following \cite{Ashtekar:2007em} and choose a factor-ordering similar to the one suggested in \cite{MartinBenito:2009aj},
\begin{align}
\wh{\mC_H} = \, & \f{3 \pi G \hbar^2}{32} \sqrt{|\nu|} (\mN_+ - \mN_-) |\nu|
(\mN_+ - \mN_-) \sqrt{|\nu|} \nn \\ &
- \f{3 \lo}{16} (2 \pi G \hbar \la_m)^{2/3} |\nu|^{5/6} (\mN_+ - \mN_-) |\nu|^{5/6} \nn \\ &
\label{ch}
- \f{\hbar^2}{2} \partial_\phi^2,
\end{align}
where the hats on operators have been dropped in order to simplify the notation, and
\begin{align}
\mN_\pm | \nu \ket :=& \, \f{1}{2} \left[ e^{\pm \la_m c / \sqrt{p}} \, \ve + \ve \, e^{\pm \la_m c / \sqrt{p}} \right] | \nu \ket \nn \\
\label{def-N}
=& \, \f{1}{2} \big[\sgn(\nu) + \sgn(\nu \pm 2) \big] | \nu \pm 2 \ket.
\end{align}
The factor-ordering choices made in \eqref{ch} and \eqref{def-N} are very similar to those previously made for the closed FLRW space-time in standard LQC in \cite{FernandezMendez:2012vi} and significantly simplify the action of the Hamiltonian constraint operator, primarily because the operator $\mN_\pm$ annihilates any eigenket $|\nu\ket$ that would otherwise change sign.  To be precise, $\mN_+$ annihilates all eigenstates of $\h \nu$ with an eigenvalue in the range%
\footnote{Note that while the states $|\nu=0\ket$ and $|\nu=-2\ket$ are not annihilated by $\mN_+$, these states are annihilated by one of the operators $\nu$ and $\sqrt{|\nu|}$ that precede or follow $\mN_+$ in $\Theta$.  Thus, when the presence of these additional operators is included in the analysis, all states $|\nu\ket$ with eigenvalues in the range $[-2, 0]$ are annihilated.  An analogous statement holds for the range of eigenstates of $\h \nu$ that are annihilated by $\mN_-$.}
$(-2, 0)$, while $\mN_-$ annihilates all eigenstates of $\h \nu$ with an eigenvalue in the range $(0, 2)$.  Because of this effect, $\ve = \sgn(\nu)$ commutes with $\mN_\pm$; and therefore the two $\ve$ operators that would otherwise have remained in each of the terms on the first two lines of \eqref{ch} commute with all of the operators there and cancel as $\sgn(\nu)^2 = 1$.

Demanding that $\wh{\mC_H}$ annihilate wave functions $\Psi(\nu, \phi)$ in the physical Hilbert space gives
\be \label{theta}
- \hbar^2 \partial_\phi^2 \Psi(\nu, \phi) = \Theta \, \Psi(\nu, \phi),
\ee
where $\Theta$, being the first two lines of \eqref{ch} multiplied by -2, acts only on $H_g$.  It is then clear that $\phi$ can play the role of a relational clock, with $\sqrt{\Theta}$ being a true Hamiltonian (assuming that $\Theta$ is essentially self-adjoint and has a positive spectrum), and the physical Hilbert space is given by the positive frequency solutions to \eqref{theta}, namely
\be
- i \hbar \, \partial_\phi \Psi(\nu, \phi) = \sqrt\Theta \, \Psi(\nu, \phi).
\ee
Then, if one is given an initial state with respect to the relational clock $\phi$, $\psi_o(\nu) = \Psi(\nu, \phi_o)$, its evolution (again with respect to the relational clock $\phi$) is
\be \label{schrodinger}
\Psi(\nu, \phi) = e^{i \sqrt\Theta (\phi - \phi_o) / \hbar} \, \psi_o(\nu).
\ee

Due to the form of $\mN_\pm$, the action of $\Theta$ on $\Psi(\nu,\phi)$ splits nicely into three parts, one part for $\nu > 0$, another for $\nu < 0$, and a final part for $\nu=0$.  It is immediately obvious that the state $|\nu=0\ket$ is annihilated by $\Theta$,
\be \label{sing}
\Theta |\nu = 0\ket = 0,
\ee
and that it is therefore a stationary state with respect to the relational clock $\phi$.  Furthermore, it is also clear that there is no eigenket $|\nu\ket$ with $\nu \neq 0$ that under the action of the Hamiltonian constraint operator is mapped to $|\nu=0\ket$.  While the states $|\nu=\pm4\ket$ and $|\nu = \pm 2\ket$ are shifted (in part) to the zero volume state, as can be seen in \eqref{ch} the new zero volume states will be acted on and annihilated by the operator $\h \nu$ (raised to some positive power).

It is in this sense that the cosmological singularity is resolved in this model: if an initial state $\psi_o(\nu)$ is non-singular (i.e., it has no support on the `singular' state $|\nu=0\ket$ which corresponds to the points in phase space of $a(t) = 0$ where the big-bang and big-crunch singularities occur in the classical theory), then under the evolution of the Hamiltonian constraint operator with respect to the relational clock $\phi$ \eqref{schrodinger}, the wave function will always remain non-singular.

Also, as already mentioned, due to the factor-ordering choice made in the definition of the shift operators $\mN_\pm$, any superposition of eigenkets $|\nu\ket$ with $\nu > 0$ will, under the action of $\Theta$, continue to have only support on $\nu>0$ since the combination of the $\sgn(\nu)$ operators will annihilate any state that would be shifted to a negative value of $\nu$, as seen in \eqref{def-N}.  For the same reason, any superposition of eigenkets $|\nu\ket$ with $\nu < 0$ will continue to have support only on $\nu < 0$ when acted upon by $\Theta$.  Thus it follows that the positive and negative sectors of $\nu$ also decouple under the action of the Hamiltonian constraint operator.

It is easy to verify that under the parity transformation \eqref{parity}, $\Pi \, \Psi(\nu, \phi) = \Psi(-\nu, \phi)$ and the shift operators $\mN_\pm$ transform as
\be
\Pi \, \mN_\pm \, \Pi = -\mN_\mp.
\ee
It immediately follows that $\Theta$ is invariant under a parity transformation,
\be
\Pi \, \Theta \, \Pi = \Theta,
\ee
and that any state that initially satisfies \eqref{parity} will continue to do so under the evolution (with respect to $\phi$) generated by the Hamiltonian constraint operator.  Due to these properties of $\Theta$, it is enough to calculate the action of this operator on the restriction of $\Psi(\nu, \phi)$ to strictly positive $\nu$, given by%
\footnote{Note that $\mN_+ |\nu\ket = |\nu+2\ket$ (for positive $\nu$ when $\sgn \, \nu = 1$) implies $\mN_+ \Psi(\nu) = \Psi(\nu - 2)$.}
\begin{align} \label{theta+}
\Theta \, \Psi(\nu) = \, &
- \f{3 \pi G \hbar^2}{16} \sqrt \nu \Big[ (\nu+2)\sqrt{\nu+4} \, \Psi(\nu+4) \nn \\ & \quad
- \sqrt{\nu} \Big(\nu + 2 + \theta_{\nu-2} (\nu - 2)\Big) \Psi(\nu) \nn \\ & \quad
+ \theta_{\nu-4} (\nu - 2) \sqrt{\nu-4} \, \Psi(\nu-4) \Big] \nn \\ &
+ \f{3 \lo}{8} (2 \pi G \hbar \la_m)^{2/3} \Big[ \Big(\nu(\nu+2)\Big)^{5/6} \Psi(\nu+2) \nn \\ & \quad
- \theta_{\nu-2} \Big(\nu(\nu-2)\Big)^{5/6} \Psi(\nu-2) \Big],
\end{align}
where, since $\Theta$ acts only on $H_g$, only the dependence of the wave function on $\nu$ is written explicitly in order to lighten the notation, and $\theta_x$ is the Heaviside function equal to 1 for positive $x$ and zero otherwise.  From \eqref{theta+} it is possible to determine the action of $\Theta$ on strictly negative $\nu$ via the relation \eqref{parity} (and recall that $\Theta |\nu=0\ket = 0$).  With this explicit definition of the action of the Hamiltonian constraint operator, it is now possible to numerically solve for the quantum dynamics (with respect to $\phi$) of any initial states of interest.

As in standard LQC, a superselection in $\nu$ occurs here, where the Hamiltonian constraint operator only couples values of $\nu$ that are separated by an integer multiple of 2.  Therefore, it is sufficient to restrict our attention to a superselection lattice denoted by the parameter $\epsilon \in (0, 2]$ and given by $L_\epsilon = \{\epsilon + 2 n, n \in \mathbb{N}\}$.  (Assuming that the initial state $\psi_o(\nu)$ is non-singular, the singular state $|\nu=0\ket$ decouples under the dynamics and need not be included in $L_\epsilon$.)  Then, from the parity condition \eqref{parity}, it follows that once $\epsilon$ is chosen for positive $\nu$, the superselection lattice for negative $\nu$ must be $L_{-\epsilon} = \{-\epsilon - 2 n, n \in \mathbb{N}\}$.

Due to the presence of quantization ambiguities already in standard LQC, there exist a number of loop quantizations of the spatially closed FLRW space-time \cite{Ashtekar:2006es, Szulc:2006ep, Corichi:2011pg, FernandezMendez:2012vi, Corichi:2013usa, Singh:2013ava}.  It has been shown numerically that the differences in these various quantum theories, while extant, are very small and do not significantly change the qualitative behaviour of the space-time, at least for states where the bounce occurs at a volume large compared to the Planck volume \cite{Corichi:2011pg, MenaMarugan:2011me}.  Since self-dual LQC is very similar to standard LQC in many aspects ---particularly in that the basic operators have similar definitions, that the quantum dynamics is given by a difference equation in the volume of the spatial 3-sphere, that the singularity is resolved since singular states decouple under the action of the Hamiltonian constraint operator, and that superselection in $\nu$ occurs--- it may be reasonable to expect that the qualitative predictions of self-dual LQC will be similar to those of standard LQC, especially in light of the robustness of standard LQC with respect to quantization ambiguities.

Nonetheless, there are also a number of apparent differences between self-dual LQC and standard LQC, of which three stand out.  The first is that the Hamiltonian constraint operator has a simpler form here since the term corresponding to the spatial curvature vanishes when self-dual variables are used, the second that the shift operator is no longer norm-preserving, and the last is that the inner product between the basis vectors $|\nu\ket$ on the kinematical Hilbert space is different than in standard LQC.  These three differences between self-dual LQC and all standard LQC models are independent of the usual LQC quantization ambiguities, but their impact on the resulting quantum dynamics is not readily apparent.  In order to determine whether the already-known results of standard LQC also hold in self-dual LQC, it is important to better understand the relation between self-dual and standard LQC.

\subsection{Relation to Standard LQC}
\label{ss.rel}

The relation between self-dual and standard LQC can be considered at both the kinematical and dynamical levels.  It is clear that the kinematical Hilbert spaces of self-dual and standard LQC are very similar, and as shall be shown here they are in fact isomorphic.  On the other hand, while the form of the Hamiltonian constraints is also quite similar, there do exist some differences between the two operators that are described below.

The easiest way to compare the two kinematical Hilbert spaces is to start from the standard LQC $H_g$, where a convenient basis is given by (see, e.g., \cite{Ashtekar:2006es} for further details)
\be
|\t p\ket = e^{i \cdot 3 \t p \t c / 8 \pi \ga G \hbar}, \qquad \bra \t p_1 |\t p_2 \ket = \de_{\t p_1, \t p_2},
\ee
and the fundamental operators are
\be \label{basic-std-lqc}
\wh{\t p} |\t p\ket = \t p |\t p\ket, \qquad
\wh{e^{i \mu \t c}} |\t p\ket  = |\t p + 8 \pi \ga G \hbar \mu / 3 \ket.
\ee
Here the tildes denote the variables of standard LQC.

In the classical theory for the spatially closed FLRW space-times, the self-dual Ashtekar connection (parametrized by $c$) and the real-valued Ashtekar-Barbero connection (parameterized by $\t c$) are related via
\be
\t c = - i \ga c + (1 + i\ga) \f{\ve \lo}{2},
\ee
and also $\t p = p$.  These relations, together with $\ve p = |p|$, suggest the definition of the basis vectors $|p\ket$ for self-dual LQC via
\begin{align}
|\t p\ket &= e^{i \cdot 3 p [ -i \ga c + (1 + i\ga) \ve \lo /2 ] / 8 \pi \ga G \hbar} \nn \\
&= e^{i \cdot 3 \lo |p| / 16 \pi \ga G \hbar} |p\ket_n,
\end{align}	
and indeed, this clearly shows that, up to a phase, the basis vectors $|\t p\ket$ of standard LQC coincide with the normalized eigenkets of self-dual LQC.

Similarly, the basic operators of self-dual LQC can be related to the basic operators of standard LQC.  Clearly, $\t p = p$, and motivated by the Baker-Campbell-Hausdorff equation (for the special case when $[X, [X,Y]] = [Y, [X,Y]] = 0$)
\be
e^X e^{\la Y} e^{-X} = e^{\la(Y + [X,Y])}, \qquad {\rm with}~\la \in \mathbb{C},
\ee
and the Poisson brackets \eqref{poisson}, it is natural to define the operator $e^{\mu c}$ as
\be \label{comp-exp}
e^{i \mu \t c} =  e^{(i - \ga) \cdot k |p| / \ga} \, e^{\ga \mu c} \, e^{-(i - \ga) \cdot k |p| / \ga},
\ee
with $k = 3 \lo  / 16 \pi G \hbar$.  From this perspective, the introduction of generalized holonomies ---and particularly of the case where $\alpha$ is purely imaginary--- occurs quite naturally: standard holonomies of the Ashtekar-Barbero connection are related to generalized holonomies (with $\alpha$ purely imaginary) of the self-dual connection and not to standard holonomies of the self-dual connection.

Then, the ansatz \eqref{comp-exp}, together with the relation between $|\t p\ket$ and $|p\ket$ and the known action of \eqref{basic-std-lqc}, leads to the definition of a shift operator given by
\be \label{shift-rel}
e^{\ga\mu c} |p\ket_n = e^{k (|p + 8 \pi G \hbar \mu / 3| - |p|)} |p + 8 \pi G \hbar / 3 \ket_n,
\ee
which is identical to the shift operator given in \eqref{shift-n}, and hence also to \eqref{shift-op}.

Thus, it is clear that the kinematical Hilbert spaces of standard LQC and self-dual LQC are isomorphic and that it is easy to move from one to another.  In particular, it is possible to define the kinematical Hilbert space of self-dual LQC by starting from the standard LQC $H_g$ and using the map described here.  Note that these definitions also automatically lead to the same reality conditions defined in Sec.~\ref{ss.kin} that are implemented by the inner product \eqref{innerP-basis}, or equivalently ${}_n\bra p_1 | p_2 \ket_n = \de_{p_1, p_2}$.

However, it is important to keep in mind that this was not the procedure followed in this paper.  Instead, the kinematical Hilbert space of self-dual LQC was constructed from scratch: following LQG the basic operators were chosen to correspond to areas and generalized holonomies, and the basis states were taken to be eigenvectors of the area operator.  Furthermore, these basis states were required to be normalizable (which meant that only generalized holonomies with purely imaginary $\alpha$ are well-defined), once again motivated by LQG.  This procedure, while it leads to the same kinematical Hilbert space for self-dual LQC as the one constructed in this section coming from standard LQC, is quite different.  The fact that both procedures give the same kinematical Hilbert space shows that the results for self-dual LQC presented here are quite robust, and indicates that there may exist different equivalent approaches to implementing the reality conditions in more general contexts as well.

In order to compare the Hamiltonian constraint operator of standard LQC and self-dual LQC, it is necessary to choose which Hamiltonian constraint operator of standard LQC is the most relevant one.  There are a number of quantization ambiguities that arise in LQC, which can be separated into three groups: field strength operator ambiguities, inverse triad operator ambiguities, and factor-ordering ambiguities.  Given the choice of the lapse $N=|p|^{3/2}$ in this paper, it is not necessary to introduce any inverse triad operators.  Also, numerical studies have been shown that factor-ordering ambiguities only affect the quantum dynamics in a very minor fashion (and that only in the deep Planck regime) \cite{MenaMarugan:2011me}, and therefore factor-ordering ambiguities will be disregarded here.  This leaves ambiguities in the definition of the operator corresponding to the field strength, which in standard LQC can be either directly represented as a non-local operator, or in terms of a non-local connection operator \cite{Singh:2013ava}.  Since the quantization procedure followed here for self-dual LQC is to express the field-strength operator in terms of the non-local connection operator \eqref{nl-conn} ---a procedure first introduced in order to complete the loop quantization of the spatially curved Bianchi space-times \cite{Ashtekar:2009um, WilsonEwing:2010rh}--- it is appropriate here to compare the self-dual Hamiltonian constraint operator to the Hamiltonian constraint of standard LQC where the field strength operator is also constructed from a non-local connection operator, which was studied for the closed FLRW space-time in \cite{Corichi:2011pg, Corichi:2013usa}.  In these papers, a different lapse is used ---for which it is necessary to introduce inverse triad operators--- and different factor-ordering choices are made than in this paper, but it is straightforward to repeat the quantization procedure presented in \cite{Corichi:2011pg, Corichi:2013usa} with the minor changes of choosing the lapse $N=|p|^{3/2}$ and making similar factor-ordering choices as the ones made in this paper for \eqref{ch}.  The Hamiltonian constraint operator for standard LQC resulting from this procedure is
\begin{align}
\wh{\wt \mC_H} = \, & - \f{3 \pi G \hbar^2}{8 \ga^2} \sqrt{|\t\nu|} (\t\mN_+ - \t\mN_-) |\t\nu| (\t\mN_+ - \t\mN_-) \sqrt{|\t\nu|} \nn \\ &
+ \f{3 \lo}{8 \ga^2} (2 \pi G \hbar \la_m)^{2/3} |\t \nu|^{5/6} (\t\mN_+ - \t\mN_-) |\t\nu|^{5/6} \nn \\ &
\label{ch-std}
- (1+\ga^2) \f{(2 \pi G \hbar^4 \la^4)^{1/3} \lo^2}{32 \ga^2} |\t\nu|^{4/3}
- \f{\hbar^2}{2} \partial_\phi^2.
\end{align}

In the attempt to translate the Hamiltonian constraint operator of standard LQC \eqref{ch-std} to the one of self-dual LQC, the key step is to replace the `improved dynamics' shift operator using the relation
\be \label{comp-N}
\t \mN_\pm = e^{(i - \ga) \cdot k |p| / \ga} \, \mN_\pm \, e^{-(i - \ga) \cdot k |p| / \ga},
\ee
which follows from \eqref{shift-rel}.  Since $\t \nu = \nu$, any differences between the two Hamiltonian constraint operators must come from differences in the shift operators, or directly from the form of the Hamiltonian constraint operator itself.

Through this relation and $\t \nu = \nu$, it is possible to rewrite the standard LQC Hamiltonian constraint operator in terms of the operators of self-dual LQC.  The result is almost exactly \eqref{ch}, with two important differences: first, all $\mN_\pm$ operators are replaced by \eqref{comp-N}, and second, there is an extra term corresponding to the spatial curvature coming from the $(1+\ga^2) E^a_j E^b_k \ep^{jk}{}_l \Omega_{ab}{}^l / \sqrt{q}$ term in the Hamiltonian constraint when expressed in terms of the Ashtekar-Barbero connection variables.

Concerning the first difference, it is clear that the operators $\mN_\pm$ and those of the type \eqref{comp-N} act differently: the $\mN_\pm$ are not norm-preserving operators while the \eqref{comp-N} operators are.  Thus, it follows that two otherwise identical Hamiltonian constraint operators where one contains the shift operators $\mN_\pm$ and the other $\t\mN_\pm$ are not equivalent.

The second difference is more obvious, since the Hamiltonian constraint expressed in terms of self-dual variables does not contain a term corresponding to the spatial curvature, unlike the case when the Hamiltonian constraint is expressed in terms of the Ashtekar-Barbero variables.  Note that these two differences do not cancel each other out (at least, not in the deep Planck regime), and thus the Hamiltonian constraint operators \eqref{ch} and \eqref{ch-std}, of self-dual LQC and of the connection-based quantization of standard LQC respectively, are truly different even for identical choices of factor-ordering and of the lapse.

Note also that it is not possible to remove the spatial curvature term by setting $\ga=i$, since the Hamiltonian constraint operator of standard LQC (which was the starting point of this construction) is only well-defined for real-valued $\ga$.  Another way to see that choosing $\ga=i$ is impossible is via the relation \eqref{comp-N} where $\ga=i$ would give an operator $\mN_\pm$ that would be norm-preserving, which is clearly in contradiction with earlier results.

While in the classical theory replacing $\t c$ and $\t p$ by $c$ and $p$ is sufficient to obtain the self-dual Hamiltonian constraint from the Hamiltonian constraint in terms of Ashtekar-Barbero variables, as seen here this simple procedure does not give the same result after a loop quantization has been performed.  So, although the kinematical Hilbert spaces of self-dual and standard LQC are isomorphic, their Hamiltonian constraint operators ---and hence their physical Hilbert spaces--- are inequivalent.

Despite the inequivalence of the Hamiltonian constraint operators of self-dual and standard LQC, their forms are very similar and differ by a quantization ambiguity related to the choice of the field strength operator: in self-dual LQC, it is the field strength corresponding to the self-dual Ashtekar connection that is promoted to be the field strength operator in the Hamiltonian constraint operator, while in standard LQC it is the field strength of the Ashtekar-Barbero connection that appears in \eqref{ch-std}.

As mentioned above, another ambiguity in the choice of the field strength operator ---namely, whether a non-local field strength operator is defined in terms of a holonomy around a closed loop or instead via a non-local connection operator defined in terms of open holonomies--- has already been studied in some detail in LQC, and it has been found that the qualitative predictions of the theory are insensitive to this ambiguity \cite{Corichi:2011pg, Corichi:2013usa, Singh:2013ava}.  (Note however that some quantitative predictions do in fact depend on the details of the loop quantization.)  This robustness of the main qualitative results of LQC, in the sense that they are independent of these quantization ambiguities%
\footnote{While the quantization ambiguity between the choice of a non-local self-dual connection operator and a non-local Ashtekar-Barbero connection operator is not the exactly the same as the ambiguity between the different definitions of the field strength operator for the Ashtekar-Barbero connection studied in \cite{Corichi:2011pg, Corichi:2013usa, Singh:2013ava}, this is in the same general family of quantization ambiguities ---i.e., ambiguities concerning the definition of the field strength operator--- and so it appears reasonable to expect the qualitative results of LQC to also be robust with respect to this quantization ambiguity.},
suggests that these key features will also appear in self-dual LQC.  In particular, the most important result of LQC is the fact that the big-bang and big-crunch singularities are resolved and replaced by a bounce, even for states that have no semi-classical interpretation, and the arguments given here suggest that this result will hold for self-dual LQC also.  Further strong evidence for the presence of a bounce in self-dual LQC is given by the effective equations presented in the following section that show that a bounce occurs for all states that are sharply-peaked and whose volume always remains much larger than $\lp^3$.

Most tellingly, however, is the fact that the Hamiltonian constraint operator for a flat FLRW space-time [obtained from \eqref{ch} by setting $\lo = 0$] is essentially identical%
\footnote{The form is identical, and the only difference is that the numerical factor $\la_m$ here is replaced by $\gamma \sqrt\Delta$ in standard LQC, where $\Delta \sim \lp^2$ is the area gap of standard LQG.}
to the Hamiltonian constraint operator for the standard loop quantum cosmology of a flat FLRW space-time, for the factor-ordering prescription%
\footnote{This particular equivalence with the `sMMO' Hamiltonian constraint operator arises due to the choice of the lapse and the factor-ordering choices made in the definition of \eqref{ch}, motivated by choices made in standard LQC in \cite{Ashtekar:2007em, MartinBenito:2009aj, FernandezMendez:2012vi}.  By making different choices for the lapse and factor-ordering, it would be possible to obtain a Hamiltonian constraint operator essentially identical to any standard LQC Hamiltonian constraint operator for the flat FLRW space-time found in the literature.}
called `sMMO' in \cite{MenaMarugan:2011me}.  The `sMMO' model has been studied in some detail, and it has been shown that the big-bang singularity is resolved and is generically replaced by a bounce --- even for states that have no semi-classical interpretation \cite{MartinBenito:2009aj, MenaMarugan:2011me} (and similar results concerning the genericity of the bounce hold for other choices of lapse and factor-ordering in standard LQC).  Thus, in the limit of vanishing spatial curvature, it is clear that a bounce occurs in self-dual LQC, and that this bounce has all of the same properties as the `sMMO' prescription of standard LQC.

An important point here is that, since spatial curvature is typically negligible in the high-curvature regime of isotropic cosmology, it is reasonable to expect that a bounce will occur for generic solutions to \eqref{ch}, including for $\lo \neq 0$, as the contributions from the spatial curvature are typically expected to be negligible when the space-time curvature is large and then the quantum dynamics will be well-approximated by those generated by the spatially flat self-dual Hamiltonian constraint operator where it is known that a bounce occurs generically, even for states that are never semi-classical.  Nonetheless, it is necessary to study the full quantum dynamics in further detail in order to confirm this expectation.

To conclude this section, although the Hamiltonian constraint operators of self-dual and standard LQC are not identical, the close relation between these theories strongly suggests that the qualitative results of standard LQC are also likely to hold in self-dual LQC, with the most important of these results being the generic appearance of a bounce even for states with no semi-classical interpretation.  However, there do exist some differences for spatially closed FLRW space-times related to the definition of the shift operator in the Hamiltonian constraint operator and also to the presence of a term corresponding to the spatial curvature.  These small differences, while not expected to strongly affect the qualitative features of the dynamics, may lead to quantitative differences between self-dual and standard LQC and it would be interesting to study these in detail in future work.

\section{The Effective Theory}
\label{s.eff}

It is by now well known that in standard LQC, for physical states that are sharply peaked at some initial `time' $\phi_o$, there exist effective equations that provide an excellent approximation to the full quantum dynamics at all times \cite{Ashtekar:2006es, Taveras:2008ke}.

These effective equations are extremely reliable because quantum fluctuations (if initially small) do not grow significantly as the high-curvature regime is approached so long as the volume of the space-time remains large compared to the Planck volume.  The reason for this is that in minisuperspace models, it is global quantities such as the total volume that are of interest and if $V \gg \lp^3$ these correspond to heavy degrees of freedom where quantum fluctuations are negligible \cite{Rovelli:2013zaa}.

These results concerning the reliability of the effective equations, originally obtained for standard LQC, are easily extended to self-dual LQC and therefore the effective equations will provide an excellent approximation to the full dynamics of sharply-peaked states in this setting also, so long as the volume $V$ remains much larger than $\lp^3$.  Thus, in this section any terms of the order $\lp^3/V$ can be dropped as these terms are negligible in the regime where the effective equations are reliable.  If one wishes to study the dynamics of a space-time that reaches a volume comparable to the Planck volume, then a full quantum evolution using the Hamiltonian constraint operator is necessary.

The effective equations are obtained by taking the Hamiltonian constraint operator \eqref{ch} and replacing all of the operators by the appropriate functions on the original phase space of the classical theory.  Then the standard Poisson relations $\d \mO = \{\mO, C_H^{eff}\}$ determine the dynamics.  For $N=1$, the effective Hamiltonian constraint is
\begin{align}
C_H^{eff} = \, & \f{3 p^{3/2}}{8 \pi G \la_m^2}
\left[ \sinh^2 \f{\la_m c}{\sqrt{p}}
- \f{\la_m \lo}{\sqrt{p}} \sinh \f{\la_m c}{\sqrt{p}} \right] \nn \\ &
\label{c-eff}
\quad + p^{3/2} \rho \approx 0;
\end{align}
since in the classical and effective theories it is possible to choose a specific orientation, $p$ has been taken to be positive.  Note that it is easy to allow for any perfect fluid simply by replacing $p_\phi^2 / 2 p^{3/2}$ by $p^{3/2} \rho$, with $\rho$ being the energy density of the perfect fluid.

This effective Hamiltonian constraint is different from the one obtained in standard LQC since there is no term corresponding to the spatial curvature, and also because the $\sin$ terms are replaced by $\sinh$, although this last change is offset by the $i$ appearing in the Poisson brackets of the self-dual variables.  Another important point is that this effective Hamiltonian constraint follows from a quantum theory where a non-local connection operator (rather than a non-local field strength operator) plays a key role in the definition of the Hamiltonian constraint operator.  This quantization choice leaves traces in the effective theory, as can be seen here by the fact that \eqref{c-eff} can be directly obtained from the classical Hamiltonian constraint \eqref{ham-cl} by the polymerization $c \to (\sqrt{p}/\la_m) \sinh(\la_m c / \sqrt{p})$, which is not the case when a non-local curvature operator is used \cite{Ashtekar:2006es}.  A more detailed discussion in the context of standard LQC on the difference between effective dynamics coming from quantum theories based on non-local connection operators versus non-local field strength operators can be found in \cite{Corichi:2011pg, Corichi:2013usa, Singh:2013ava}.

Recalling the fundamental Poisson brackets given in \eqref{poisson}, one of the effective equations of motion is
\be
\d p = \! - \f{2ip}{\la_m} \left[ \sinh \f{\la_m c}{\sqrt{p}} \cosh \f{\la_m c}{\sqrt{p}}
- \f{\la_m \lo}{2 \sqrt{p}} \cosh \f{\la_m c}{\sqrt{p}} \right],
\ee
which, after using the effective scalar constraint \eqref{c-eff} twice, gives the modified Friedmann equation
\be \label{hubble-eff}
H^2 = \left(\f{\d p}{2 p} \right)^2 = \left( \f{8 \pi G}{3} \rho - \f{\lo^2}{4 p} \right) \times
\left( 1 - \f{\rho}{\rho_c} \right),
\ee
where $\rho_c = 3 / 8 \pi G \la_m^2$, and terms of the order $\la_m / \sqrt{p}$ have been dropped since the derivation of these equations makes the assumption that the volume of the space-time always remains much larger than the Planck volume, and it is also assumed that the discreteness scale $\la_m$ is of the order of the Planck length.

Note that while this is not the same effective Friedmann equation as the one derived in \cite{Ashtekar:2006es} (although the qualitative aspects of the effective dynamics are unchanged), it is in fact exactly the effective Friedmann equation found in \cite{Corichi:2011pg} for the connection-based standard loop quantization, at least up to terms of the order $\la_m / \sqrt p$ which are assumed to be negligible.  This provides further evidence of the close relation between self-dual LQC and the connection-based loop quantization using Ashtekar-Barbero variables.  (Note that the differences between the Hamiltonian constraint operators of these theories described in Sec.~\ref{ss.rel} are encoded here in differences of the order $\la_m / \sqrt p$ that are negligible in the regime of validity of the effective equations.)

Since there are only corrections to the gravitational sector of the Hamiltonian constraint in the effective theory, the continuity equation is unchanged,
\be
\d \rho + 3 H (\rho + P) = 0.
\ee
Then, taking the time derivative of \eqref{hubble-eff} and using the continuity equation gives
\begin{align}
\f{\ddot{a}}{a} = \, & \f{8 \pi G}{3} \rho \left(1 - \f{\rho}{\rho_c} \right)
- 4 \pi G (\rho + P) \left(1 - \f{2\rho}{\rho_c} \right) \nn \\ &
\label{raychaudhuri-eff}
\quad - \f{3 \lo^2}{8 p} \left( \f{\rho + P}{\rho_c} \right).
\end{align}
Clearly, the classical Friedmann equations are recovered in the limit $\rho_c \to \infty$.

The effective Friedmann equations \eqref{hubble-eff} and \eqref{raychaudhuri-eff} show that a bounce occurs when the energy density of the matter field equals the critical energy density $\rho_c$, and therefore, for sharply-peaked states, the big-bang and big-crunch singularities that generically arise in the FLRW solutions to general relativity are automatically resolved in self-dual loop quantum cosmology and are replaced by a bounce, just as in standard loop quantum cosmology.

Finally, setting $\lo=0$ gives the effective equations for the spatially flat FLRW space-time, which are identical to the effective equations in standard LQC for that space-time \cite{Ashtekar:2006wn}.

\section{Discussion}
\label{s.disc}

In this paper, it was shown how a loop quantum cosmology for spatially closed FLRW space-times can be constructed starting from self-dual variables.  The resulting quantum theory is closely related to standard LQC in that: (i) the kinematical Hilbert space is the same, (ii) the dynamics are given by a difference equation under which the big-bang and big-crunch singularities are resolved since the non-singular states decouple from the singular states under the action of the Hamiltonian constraint operator, and (iii) for sharply peaked states (and where $\bra |V| \ket$ is always much larger than the Planck volume) the full quantum dynamics are well-approximated by effective equations which show that the classical singularity is replaced by a non-singular bounce due to quantum gravity effects.  Note however that there are some differences between self-dual LQC and standard LQC in the explicit expressions of the Hamiltonian constraint operator which would be interesting to study in future work.  Finally, in the limit of a spatially flat space-time (obtained by setting $\lo = 0$ in the relevant equations), one recovers exactly the results of standard LQC for flat FLRW space-times (with the same factor-ordering ambiguities) that have been studied in considerable depth in the literature.

It is interesting to compare the results obtained here with those of \cite{Achour:2014rja} where a different procedure is followed in order to determine the form LQC takes when the Barbero-Immirzi parameter is set to $\gamma = i$.  This other procedure is to take the results of standard LQC and to simultaneously (i) perform an analytic continuation $\gamma \to i$ and (ii) change from calculating the holonomy in the $j=1/2$ representation of SU(2) to the lowest non-trivial continuous representation of SU(1,1).  It is then found that the resulting theory is different from what is obtained here.  Most strikingly, while the effective equations show that the big-bang singularity is also replaced by a bounce, in \cite{Achour:2014rja} the bounce is found to be necessarily strongly asymmetric.  The differences between these two approaches probably arise due to the fact that in this paper the generalized holonomies were taken to be in the $j=1/2$ representation of SU(2), just as in standard LQC.  This suggests that while using self-dual variables does not strongly affect the qualitative features of LQC, representing the holonomies in terms of the continuous representation of SU(1,1) does lead to significant changes in the resulting theory.

One of the two key new steps that are necessary in order to define self-dual LQC as presented here is the implementation of the reality conditions in the kinematical Hilbert space by choosing an appropriate inner product.  Due to the isotropy of the space-time, the spin-connection is almost independent of the phase space variables (depending only on the orientation of the triads) and this simplification made the task of imposing the reality conditions relatively easy.  Nonetheless, the fact that this could be done for the spatially closed FLRW space-time raises the hope that it may be possible for more complicated space-times as well, and a natural next step in this direction would be to consider the Bianchi models with non-vanishing spatial curvature.  If the reality conditions can also be handled in more complex minisuperspaces like the Bianchi type IX space-time, it may be worth revisiting the possibility of using self-dual variables in full loop quantum gravity.

Another key result of this paper is that operators corresponding to standard holonomies are not well-defined in the kinematical Hilbert space.  Instead, it was necessary to introduce a family of generalized holonomies parametrized by $\alpha$ (with the standard holonomies recovered for $\alpha = 1$), and it was only for purely imaginary $\alpha$ that the generalized holonomy operators were well-defined.  This can be understood to come from the fact that it is the extrinsic curvature part of the Ashtekar connection that is conjugate to the densitized triad, and thus ---because the kinematical Hilbert space is, loosely speaking, essentially the space of square integrable functions with respect to the Haar measure on the Bohr compactification of the extrinsic curvature since the spin-connection is a constant--- it is operators corresponding to complex exponentials of the extrinsic curvature that are well-defined in the quantum theory.  This explains why only generalized holonomies with purely imaginary $\alpha$ are well-defined for self-dual LQC where $\gamma=i$, while normal holonomies (with $\alpha = 1$) are well-defined in standard LQC where $\gamma \in  \mathbb{R}$.

While it is necessary to go beyond the study of the simplest FLRW space-times in quantum cosmology before drawing any major conclusions, this result does raise the possibility that if one chooses to work with the self-dual variables, then standard holonomies ---which as pointed out above are perfectly well-defined in LQC and LQG for the real-valued Ashtekar-Barbero variables--- perhaps should not be chosen to be one of the basic variables to be promoted to be fundamental operators in the quantum theory.  Instead, for self-dual variables it may be necessary to consider a different type of operator such as the generalized holonomies with purely imaginary $\alpha$ introduced here.

\acknowledgments

I would like to thank Joseph Ben Geloun, Karim Noui, Daniele Oriti, Alejandro Perez, Jorge Pullin, Hanno Sahlmann and Casey Tomlin for helpful discussions.
This work is supported by a grant from the John Templeton Foundation.



\raggedright

\end{document}